# Trimers of MnO$_6$ octahedra and ferrimagnetism of Ba$_4$NbMn$_3$O$_{12}$

Loi T. Nguyen, Tai Kong and R.J. Cava

Department of Chemistry, Princeton University, Princeton, New Jersey 08544, USA

**Abstract**

Ba$_4$NbMn$_3$O$_{12}$ is reported, synthesized by a solid state method in air. The crystal structure, determined by performing refinements on room temperature powder X-ray diffraction data by the Rietveld method, consists of Mn$_3$O$_{12}$ trimers in the configuration of three face-sharing MnO$_6$ octahedra, with the trimers arranged in triangular planes. An effective moment of 4.82 μ$_B$/f.u is observed and competing antiferromagnetic and ferromagnetic interactions between Mn ions are inferred from the Weiss temperature of -4 K and the ferrimagnetic ordering transition of approximately 42 K. Ba$_4$NbMn$_3$O$_{12}$ is a semiconductor with a transport activation energy of 0.37 eV.

**Introduction**

Manganese-based perovskites have been intensively researched due to the metal-insulator (MI) transitions and colossal magnetoresistance (CMR)[1,2,3,4] in this family. The interesting properties have been attributed in part to the competition between a ferromagnetic metallic state and an antiferromagnetic insulating state, and the presence of charge-ordering. Manganates are also well known due to other structural, electronic, and magnetic characteristics[5,6]. In the hexagonal perovskite $RMnO_3$ phases, for example, the electric polarization is reported to be the result of a structural transition[7] and the improper nature of ferroelectricity to a network of coupled structural and magnetic vortices that induce domain wall magnetoelectricity and magnetization[8].

Of particular interest for CMR are materials with lower structural dimension based on the Ruddlesden-Popper series[9], but quasi-1D materials are also of interest[10]. Instead of sharing corners in a 3D network as in corner-sharing perovskites, the $MnO_6$ octahedra in the quasi-1D materials can share faces with each other to form $Mn_2O_9$ dimers or $Mn_3O_{12}$ trimers. These Mn mini-clusters are isolated from other clusters in the structure. This type of $MnO_6$ configuration is relatively uncommon because the Mn-Mn distances of about 2.5 Å within the clusters are shorter than the 3.5 Å typically found in corner-sharing geometries[11]. Oxygen deficient $BaMnO_{3-x}$ and $(Ba,Sr)MnO_{3-x}$ phases are good examples of materials where clusters of face-sharing $MnO_6$ octahedra are found[12].

In the current work, we report the synthesis and initial characterization of the mixed valent $Mn^{3+}/Mn^{4+}$ compound $Ba_4NbMn_3O_{12}$. The crystal structure of this previously unreported phase, which is much like that of $Ba_4REMn_3O_{12}$ ($RE$ = Ce, Pr)[13], consists of face-sharing $MnO_6$ octahedra forming $Mn_3O_{12}$ trimers with their long axes aligned along the hexagonal $c$ axis of the trigonal symmetry crystallographic cell. From magnetic susceptibility measurements, the effective magnetic moment is found to be 4.82 µ$_B$/f.u., consistent with a simple spin configuration within the trimers. Resistivity measurements show that Ba$_4$NbMn$_3$O$_{12}$ is semiconducting with a transport gap of 0.37 eV. The current phase is chemically and structurally distinct from recently reported Ba$_8$MnNb$_6$O$_{24}$, which, among other differences, does not appear to contain any face shared octahedra[14].

**Methods**

Polycrystalline samples of Ba$_4$NbMn$_3$O$_{12}$ were synthesized by solid-state reaction using BaCO$_3$, MnO$_2$, and Nb$_2$O$_5$ (Alfa Aesar, 99.9%, 99.999%, and 99.5%, respectively) as starting materials. Reagents were mixed thoroughly in the appropriate ratio, placed in an alumina crucible, and heated in air at 900 °C for 24 hours. The resulting powder was re-ground, pressed into a pellet and heated in air at 1100 °C for 24 hours, and then at 1300 °C for 12 hours. The phase purity and crystal structure were determined through powder X-ray diffraction (PXRD) using a Bruker D8 Advance Eco with Cu Kα radiation and a LynxEye-XE detector. The structural refinements were performed with *GSAS*[15]. The crystal structure drawing was created by using the program *VESTA*[16].

The magnetic susceptibility of Ba$_4$NbMn$_3$O$_{12}$ powder was measured by a Quantum Design Physical Property Measurement System (PPMS) DynaCool equipped with a VSM option. The magnetic susceptibility of Ba$_4$NbMn$_3$O$_{12}$, defined as *M*/*H*, where *M* is the sample magnetization and *H* is the applied field, was measured at the field of *H* = 1 kOe from 1.8 K to 300 K; some additional measurements were performed in an applied field of 100 Oe. (Relatively low fields are employed so as to not overly disrupt the magnetic system through the measurements.) The resistivity of Ba$_4$NbMn$_3$O$_{12}$ was measured by the DC four-contact method in the temperature range 200 K to 350 K with the PPMS. The sample was pressed, sintered, and cut into pieces with the approximate size 1.0 × 2.0 × 1.0 mm$^3$. Four Pt contact wires were connected to the sample using silver paint. The specific heat was measured from 1.8 K to 200 K by a PPMS DynaCool equipped with a heat capacity option.

**Results and discussion**

The powder x-ray diffraction pattern and structural refinement of $Ba_4NbMn_3O_{12}$ are shown in **Figure 1**. The structure of $Ba_4NbRu_3O_{12}$[17] was used as the starting model. $Ba_4NbMn_3O_{12}$ crystallizes in a rhombohedral structure with the space group *R*-3*m* (No. 166). Refinements in which the Mn to Nb ratio was allowed to vary from 3:1 or Nb/Mn site mixing was permitted were not satisfactory. The lattice parameters and structural parameters for $Ba_4NbMn_3O_{12}$ are summarized in **Table 1**. The structure consists of three $MnO_6$ octahedra connected by face-sharing to form $Mn_3O_{12}$ trimers, as shown in **Figure 2**. The crystal structure of $Ba_4NbMn_3O_{12}$ is similar to those of two known rare-earth containing compounds $Ba_4REMn_3O_{12}$ (RE = Ce, Pr)[13] in the same family. The corner-sharing $NbO_6$ and $Mn_3O_{12}$ trimers in $Ba_4NbMn_3O_{12}$ alternate along *c* to generate the 12-layer hexagonal perovskite structure. Individual $Mn_3O_{12}$ trimers are corner-sharing with the non-magnetic $NbO_6$ octahedra, and not to other trimers, such that the magnetic coupling between trimers is of the Mn-O-O-Mn super-super exchange type. Within each trimer, the distance between Mn atoms is 2.4694(1) Å, quite short due to the face-sharing[11], favoring strong magnetic interactions between them. The Mn-O distances are 1.903(4) Å and 1.882(4) Å for the outer $MnO_6$ octahedra, reflecting a small distortion, and 1.895(4) Å for the inner $MnO_6$ octahedron, which has an ideal shape. These Mn-Mn and Mn-O bond lengths are in agreement with those found for the 10-layer hexagonal perovskite $Ba_5Sb_{0.64}Mn_{4.36}O_{15-\delta}$[18], for example.

The temperature-dependent magnetic susceptibility of $Ba_4NbMn_3O_{12}$ and its reciprocal are plotted in **Figure 3**. The magnetic data for $Ba_4NbMn_3O_{12}$ from 100 K to 275 K are well fit to the Curie-Weiss law $\chi = \frac{C}{T-\theta_{CW}} + \chi_o$, where $\chi_o$ is the temperature-independent part of the susceptibility, $C$ is the Curie constant, and $\theta_{CW}$ is the Curie-Weiss temperature. The least squares fitting yields $\chi_o = 2.79 \times 10^{-3}$ emu Oe$^{-1}$ mol-f.u.$^{-1}$, $\mu_{eff} = 4.82$ $\mu_B$/f.u, and $\theta_{CW} = -4$ K. The magnetic hysteresis loops for Ba$_4$NbMn$_3$O$_{12}$ from -2 to 2 T at 2 K, 25 K and 150 K are shown in **Figure S1 (See SI)**. A coercive force of 0.2 T and the remnant magnetization of 0.04 $\mu_B$/f.u. are observed at 2 K and 0.04 T and 0.02 $\mu_B$/f.u. at 25 K. The overall behavior is like that of a ferrimagnet. At 150 K, the magnetization is linearly proportional to the field and there is no hysteresis loop.

Because Ba$^{2+}$ and Nb$^{5+}$ (which has no electrons in $d$ orbitals) are non-magnetic, the magnetic properties of Ba$_4$NbMn$_3$O$_{12}$ are determined by the intertrimer and intratrimer interactions of the Mn$_3$O$_{12}$ units. Formally, one Mn$^{3+}$ and two Mn$^{4+}$ are found in each Mn$_3$O$_{12}$ trimer. Thus, in the paramagnetic state, where the moments fluctuate, the total magnetization per formula unit is expected to be the sum of the magnetizations of two Mn$^{4+}$ and one Mn$^{3+}$, which, should all the moments be free to respond to an applied field for temperatures below 300 K, yield a total effective moment of 6.16 (low spin Mn$^{3+}$) or 7.35 (high spin Mn$^{3+}$) $\mu_B$/f.u. This magnetic state is not compatible with the observed data. **Figure 4** shows, in contrast, a schematic of two simple hypothetical magnetic states within each trimer, which, to be consistent with our magnetic data, in these simple scenarios, either two Mn$^{4+}$ bearing spin-3/2 point in the same direction and opposite to the low spin (S=1) Mn$^{3+}$ between them or two Mn$^{4+}$ bearing spin-3/2 point in opposite directions and the Mn$^{3+}$ (high spin) with the spin-2 is between them. With this proposed magnetic coupling within the trimer, each trimer has spin-2 and a calculated effective moment of 4.90 $\mu_B$/f.u. This is in excellent agreement with the observed effective moment of 4.82 $\mu_B$/f.u in the susceptibility measurements, indicating that either of these trimer-based spin models is likely to successfully describe the magnetic state of Ba$_4$NbMn$_3$O$_{12}$ above the three-dimensional magnetic ordering temperature near 42 K. Compared to $Bi_3Mn_3O_{11}$[19], for example, which has mixed-valent Mn, an effective magnetic moment of 6.27 $\mu_B$/f.u and a Curie-Weiss temperature of 222 K, both the effective moment and Curie-Weiss temperature of $Ba_4NbMn_3O_{12}$ are smaller. The strong coupling of the Mn spins within the trimers to create a magnetic Mn "molecule" made of three Mn's in the current material is consistent with what has previously been found for $(Ba,Sr)MnO_3$ phases[12,20]. The fitted $\chi_0$ in this scenario is then a representation of the remnant susceptibility of the magnetic system after the local-only magnetic ordering has set in within the individual trimers at temperatures above 300 K.

**Figure 5** shows the field-cooled (FC) and zero-field-cooled (ZFC) DC susceptibility in an applied field of 100 Oe for $Ba_4NbMn_3O_{12}$. The increases of the FC and ZFC susceptibility below 42 K reconfirm the magnetic transition temperature observed in the magnetic susceptibility data shown in **Figure 3**, and the $\chi T$ vs. T data shown in the inset defines the temperature of the magnetic transition at the same temperature through the dramatic change in the Curie constant.

The resistivity of $Ba_4NbMn_3O_{12}$ is plotted as a function of reciprocal temperature in **Figure 6**. Resistivity data from 300 to 350 K were fit to the standard model $\rho = \rho_o e^{\frac{E_a}{k_b T}}$, and the transport activation energy $E_a$ was calculated to be 0.37 eV. The inset shows the increase in resistivity when cooling. With the activation energy of 0.37 eV, $Ba_4NbMn_3O_{12}$ is a semiconductor, similar to other trimer-based compounds ($Ba_4NbRu_3O_{12}$ and $Ba_4LnRu_3O_{12}$ and $Ba_4LnIr_3O_{12}$)[17,21].

**Figure 7A** shows the specific heat divided by temperature of $Ba_4NbMn_3O_{12}$ measured from 1.8 K to 200 K in its main panel, with the raw $C_p$ data shown in the inset. At 200 K, $C_p$ has not yet reached the saturation value of 3NR (N is the number of atoms), but this is often encountered in materials where different atomic masses and strong bonds between atoms lead to very high vibrational frequencies [22]. Three characteristic features are seen. The λ-anomaly corresponds to the magnetic transition around 42 K and is shown in **Figure 7B.** In this case we employed a polynomial function fit to the data above and below the anomaly to estimate the phonon background; at this level of analysis we do not attach any significance to the function employed to estimate the phonon part. In this way the magnetic entropy change in the higher temperature range is estimated to be 0.45 J mol$^{-1}$ K$^{-1}$. This relatively weak anomaly implies that there is a significant remnant of magnetic entropy in Ba$_4$NbMn$_3$O$_{12}$ below the 42 K transition. The specific heat in the lower temperature range is shown in more detail in **Figure 7C,** which shows another entropy loss at 5-6 K. In this case the phonon part of the specific heat is estimated by a Debye-Einstein model, again to which we attribute no physical significance at this level of analysis. The magnetic entropy at this lower temperature, which does not clearly yield a feature in the magnetic susceptibility, is found to be 2.67 J mol$^{-1}$ K$^{-1}$. For these two temperature regions only, only one-fourth of the magnetic entropy expected for an S=2 Heisenberg system is recovered. There is a more subtle feature in C$_p$/T at around 30 K that may hold a significant amount of entropy, but we cannot reasonably analyze it at this time. Aside from the most general conclusions described here, analysis of the specific heat cannot be considered as well-established at present, as we have not been able to make a non-magnetic analog to use to best model the phonon contribution to the specific heat of this material. A straightforward analysis (**See the SI, Figure S2 and S3**) shows that the observed magnetic transition near 42 K cannot possibly be due to the presence of an Mn$_3$O$_4$ impurity.

**Conclusion**

Ba$_4$NbMn$_3$O$_{12}$, a previously unreported material synthesized by a solid state method, crystallizes in a 12-layer hexagonal perovskite unit cell with the *R*-3*m* space group. Differences in the Mn-O bond lengths in the octahedra, although subtle, may reflect the presence of charge ordering, although we do not believe that the distinction is great enough to warrant assigning specific charge states to specific octahedra at the present time. The material has a high, temperature-dependent resistivity, indicating that it is semiconducting, with a transport activation energy of 0.37 eV measured on a sintered polycrystalline pellet. From magnetic susceptibility measurements, the Curie-Weiss temperature is calculated to be -4 K, indicating the presence of competing ferromagnetic and antiferromagnetic interactions between Mn trimers. The effective moment of 4.82 $\mu_B$/f.u agrees with the value of 4.90 $\mu_B$/f.u that can be deduced from a simple hypothetical magnetic model in which the spins within the trimer are essentially already ordered by 300 K, although the trimer-trimer ordering does not occur until much lower temperatures. The ordering of the trimer spins with respect to those in other trimers is what most likely gives rise to the magnetic transition observed around 42 K. Heat capacity measurements show a weak anomaly at around 42 K, supporting the magnetic ordering transition seen in magnetic susceptibility. Only small amounts of magnetic entropy are observed through the specific heat, consistent, in principle, with the proposal that most of the magnetic entropy is lost at higher temperature than is studied here. A non-magnetic analog for this system would be of interest to account for the phonon contribution in the specific heat so that a more detailed interpretation of the entropy will be possible. Magnetic neutron diffraction will also be of future interest to establish the nature of the magnetism in this material. As for Mn-based conventional perovskites, hexagonal Mn-based perovskites appear to display rich magnetic phenomena.

**Conflicts of interest**

There are no conflicts of interest to declare.

**Acknowledgement**

This work was supported in its entirety by the Department of Energy Division of Basic Energy Sciences, through the Institute for Quantum Matter, grant DE-FG02-08ER46544. The authors thank Daniel Khomskii for valuable discussions about manganite magnetism.

**Table 1.** Structural parameters for $Ba_4NbMn_3O_{12}$ at 300 K. Space group *R-3m* (No. 166).

| Atom | Wyckoff. | Occ. | x | y | z | $U_{iso}$ |
|---|---|---|---|---|---|---|
| Ba1 | 6c | 1 | 0 | 0 | 0.12847(4) | 0.0344(3) |
| Ba2 | 6c | 1 | 0 | 0 | 0.28592(4) | 0.0304(3) |
| Nb | 3a | 1 | 0 | 0 | 0 | 0.0083(3) |
| Mn1 | 3b | 1 | 0 | 0 | ½ | 0.0195(3) |
| Mn2 | 6c | 1 | 0 | 0 | 0.41218(4) | 0.0146(3) |
| O1 | 18h | 1 | 0.4776(6) | 0.5224(6) | 0.12239(4) | 0.0212(3) |
| O2 | 18h | 1 | 0.4908(6) | 0.5092(6) | 0.29327(4) | 0.0127(3) |

a = 5.71825(3) Å, c = 28.1158(3) Å

$R_{wp}$ = 10.30%, $R_p$ = 8.02%, $R_F^2$ = 10.80%

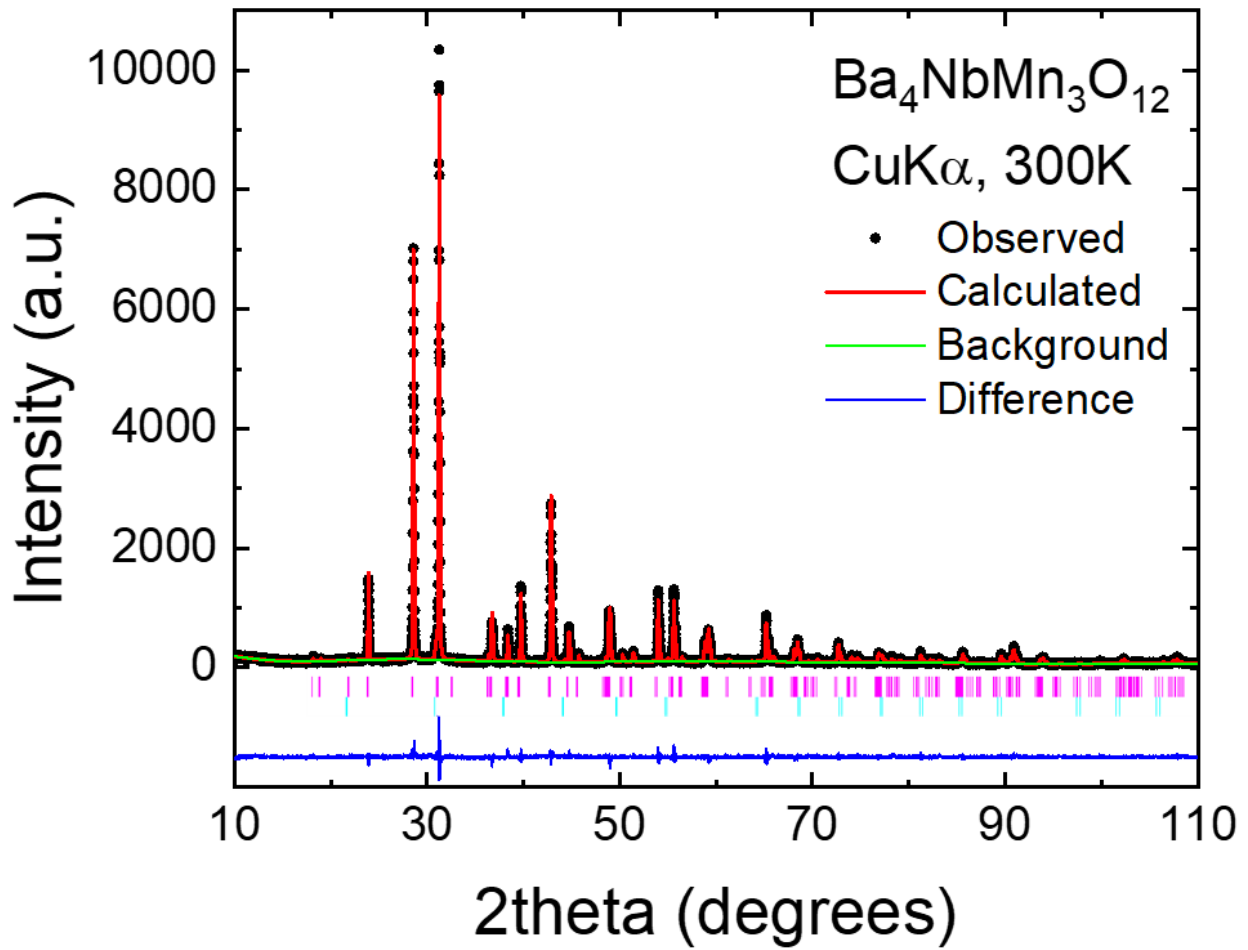

**Figure 1 (Color online):** Rietveld Powder X-ray diffraction refinement for $Ba_4NbMn_3O_{12}$ in space group *R*-3*m*. The observed X-ray pattern is shown in black, calculated in red, difference ($I_{obs}$-$I_{calc}$) in blue, the background in green, and tick marks denote allowed peak positions in pink ($Ba_4NbMn_3O_{12}$) and in cyan ($BaNb_{0.5}Mn_{0.5}O_3$).

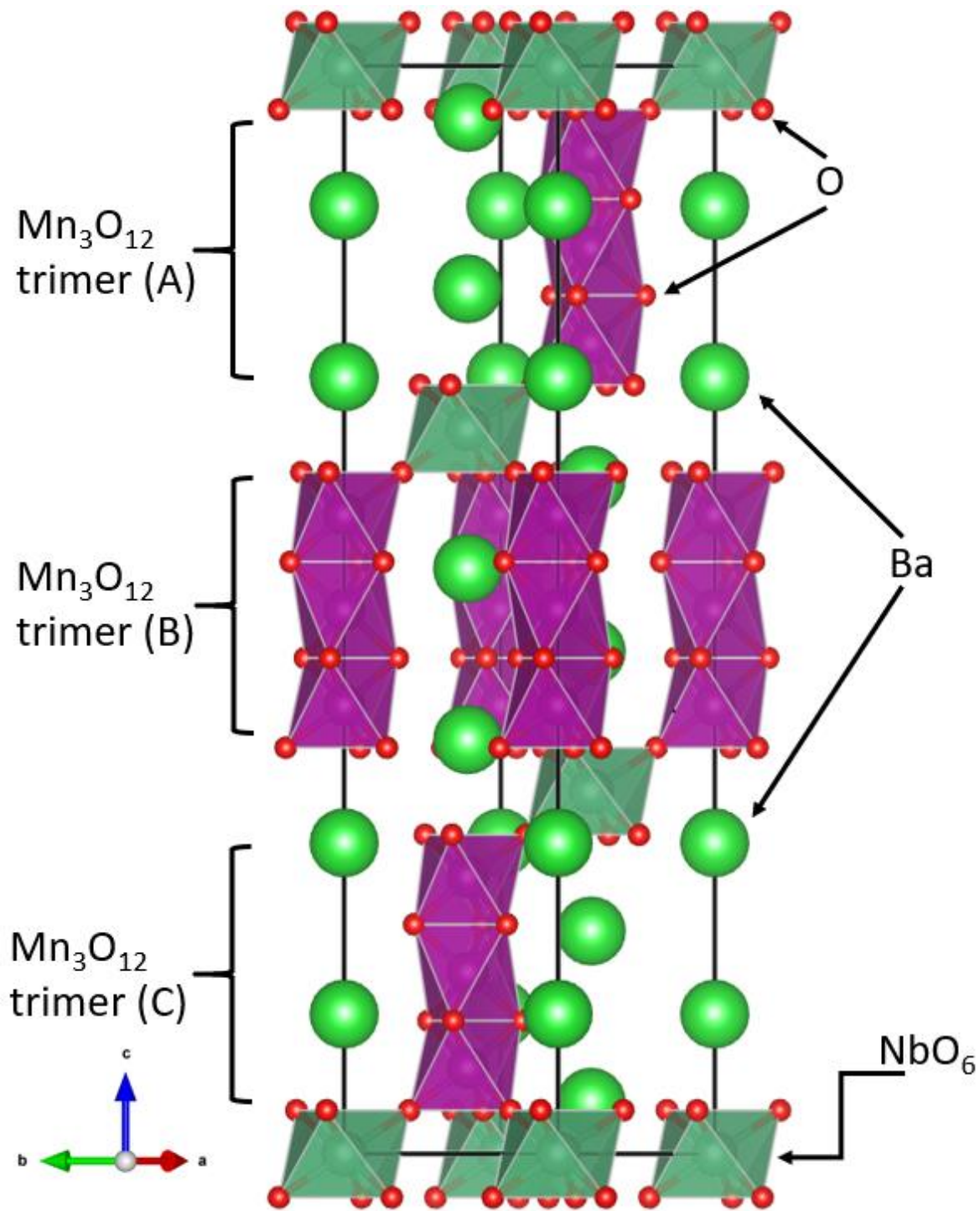

**Figure 2 (Color online): The crystal structure of Ba$_4$NbMn$_3$O$_{12}$. The NbO$_6$ octahedra (dark green) share corners with the Mn$_3$O$_{12}$ (purple) trimers made from face-shared MnO$_6$ octahedra, to form what is frequently referred to as a "12-layer" hexagonal perovskite structure. Barium is green, and oxygen is red.**

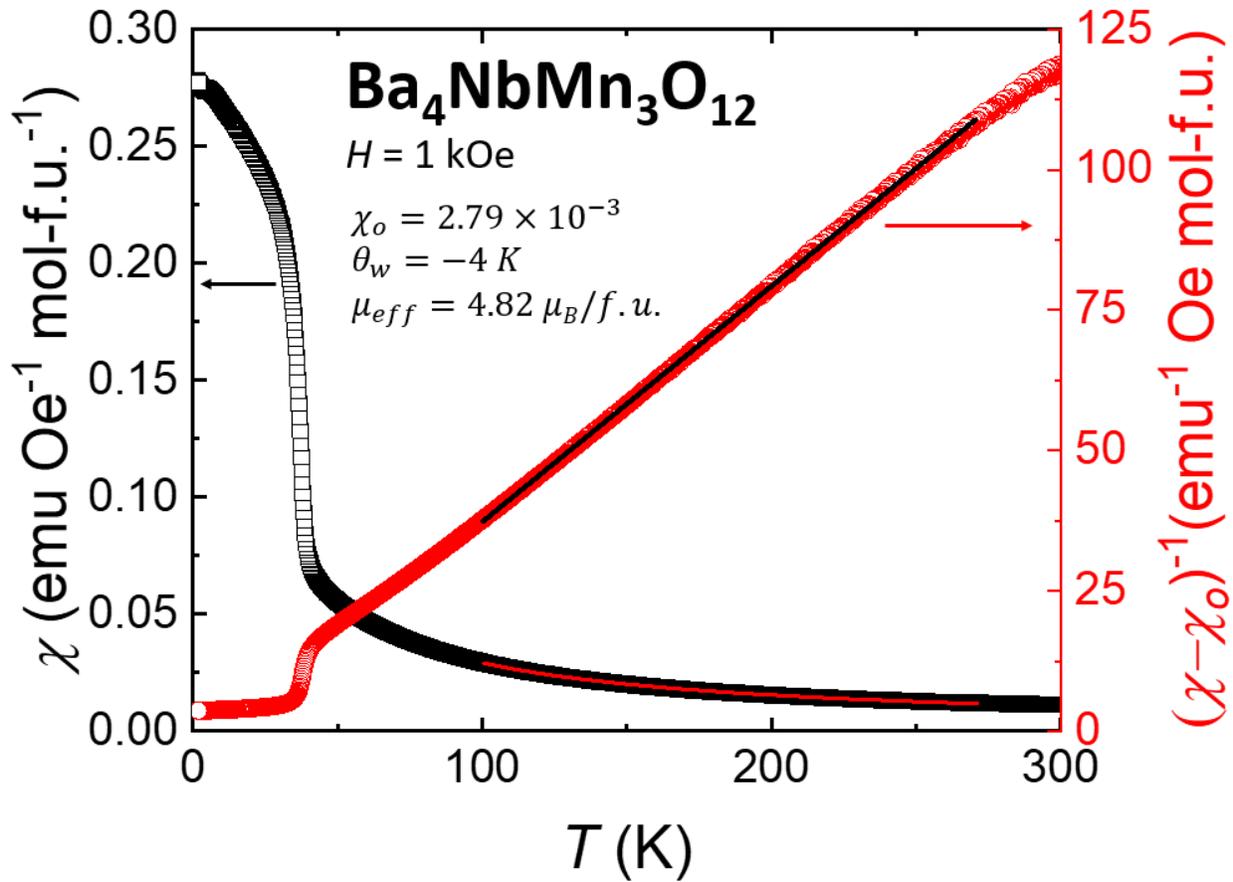

**Figure 3 (Color online):** The temperature dependence of the magnetic susceptibility and the inverse of the difference between the magnetic susceptibility and the temperature independent magnetic susceptibility ($\chi_o$ = 2.79×10$^{-3}$ emu Oe$^{-1}$ mol- f.u.$^{-1}$) for Ba$_4$NbMn$_3$O$_{12}$. The applied field was 1 kOe. The red solid line is the susceptibility fit to the Curie-Weiss law for T from 100 K-275 K.

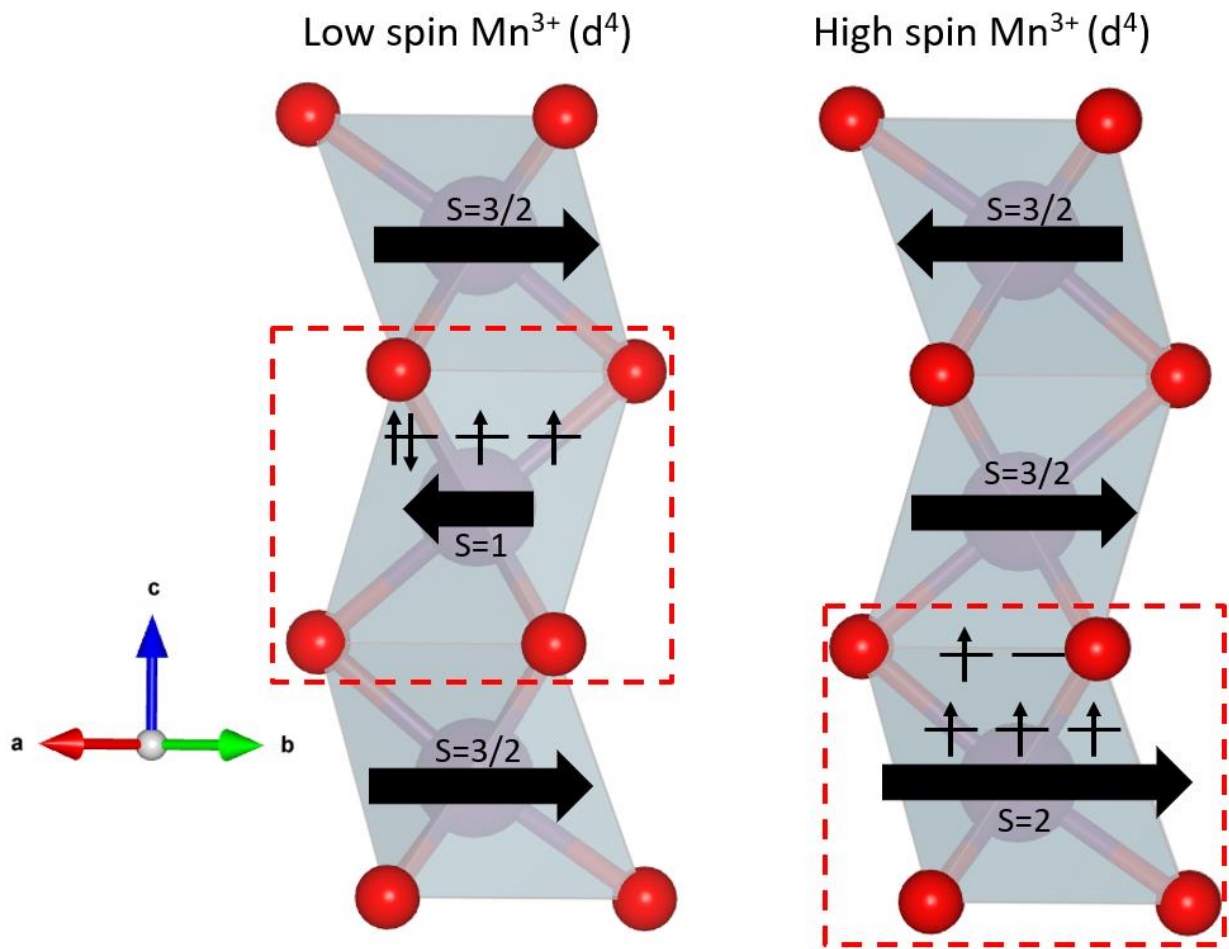

**Figure 4 (Color online): Two schematics of possible simple hypothetical arrangements for $Mn^{3+}$ (either S=1 for low spin (left model) or S=2 for high spin (right model)) and $Mn^{4+}$ (always S=3/2) in $Ba_4NbMn_3O_{12}$. Both hypothetical arrangements yield spin-2 for each $Mn_3O_{12}$ trimer, $\mu_{eff}$ = 4.90 $\mu_B$/f.u., which compares very favorably to the observed $\mu_{eff}$ = 4.82 $\mu_B$/f.u.**

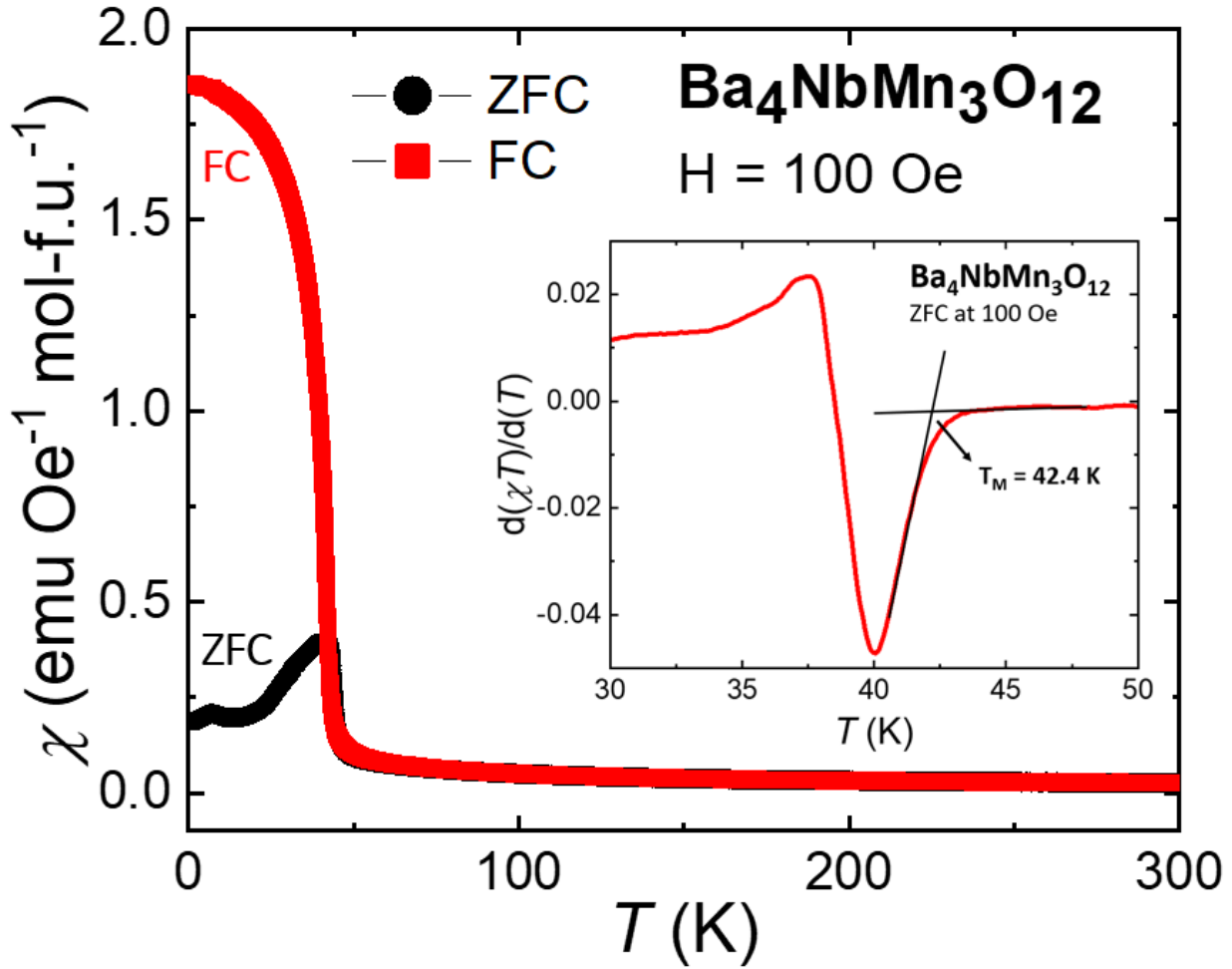

**Figure 5 (Color online):** Field Cooled (FC) and Zero Field Cooled (ZFC) DC magnetic susceptibility in an applied field of 100 Oe for $Ba_4NbMn_3O_{12}$ from 2-300 K. The inset shows the first derivative of magnetic susceptibility multiplied by temperature as function of temperature (a measure of the Curie constant). The magnetic transition temperature is found to be 42.4 K by extrapolating the derivative curves.

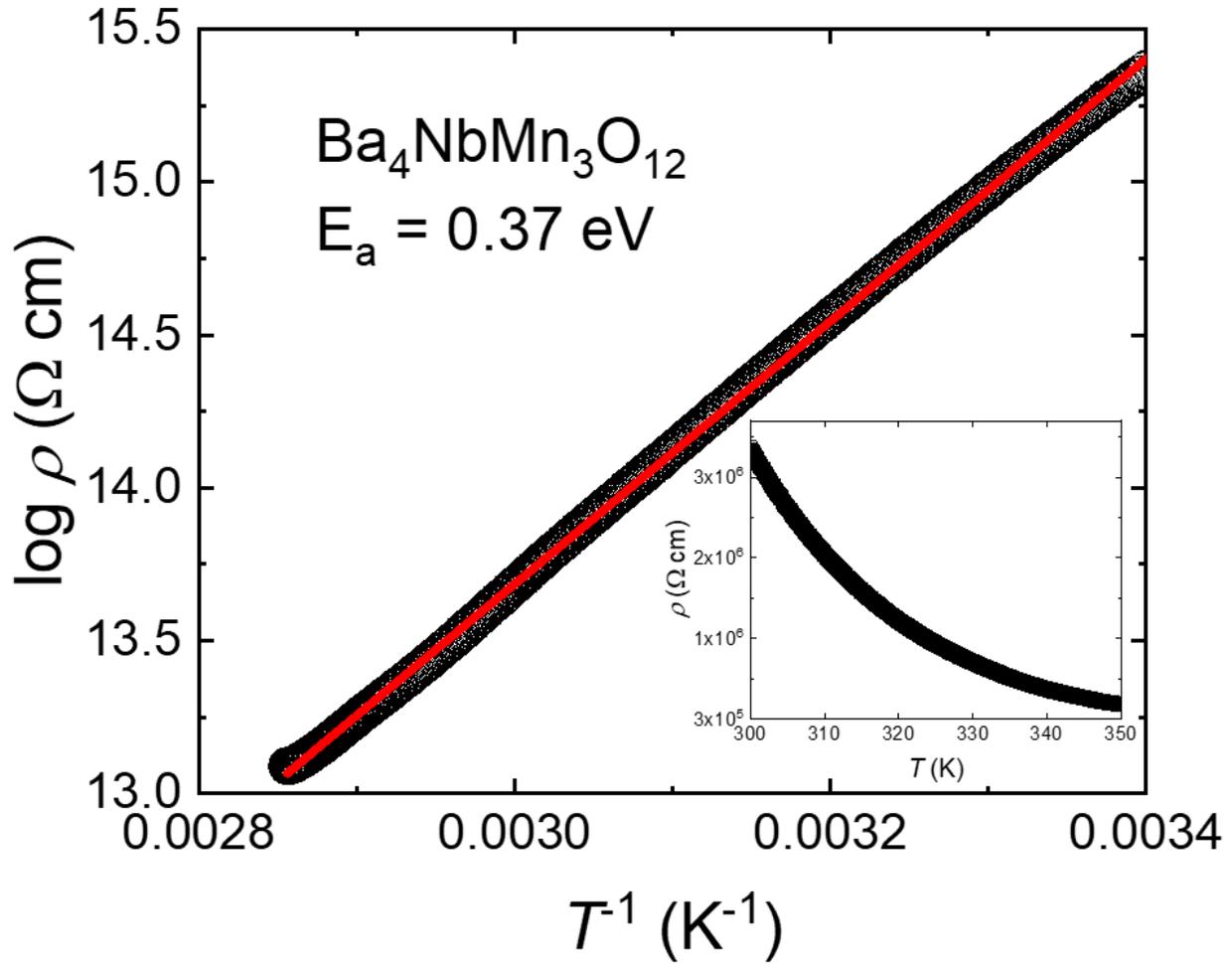

**Figure 6 (Color online):** The resistivity of a sintered polycrystalline pellet of Ba$_4$NbMn$_3$O$_{12}$ as a function of temperature (Inset) and inverse temperature (Main Panel). The data was fit to the model $\rho = \rho_o e^{\frac{E_a}{k_b T}}$ (red line) with E$_a$ = 0.37 eV.

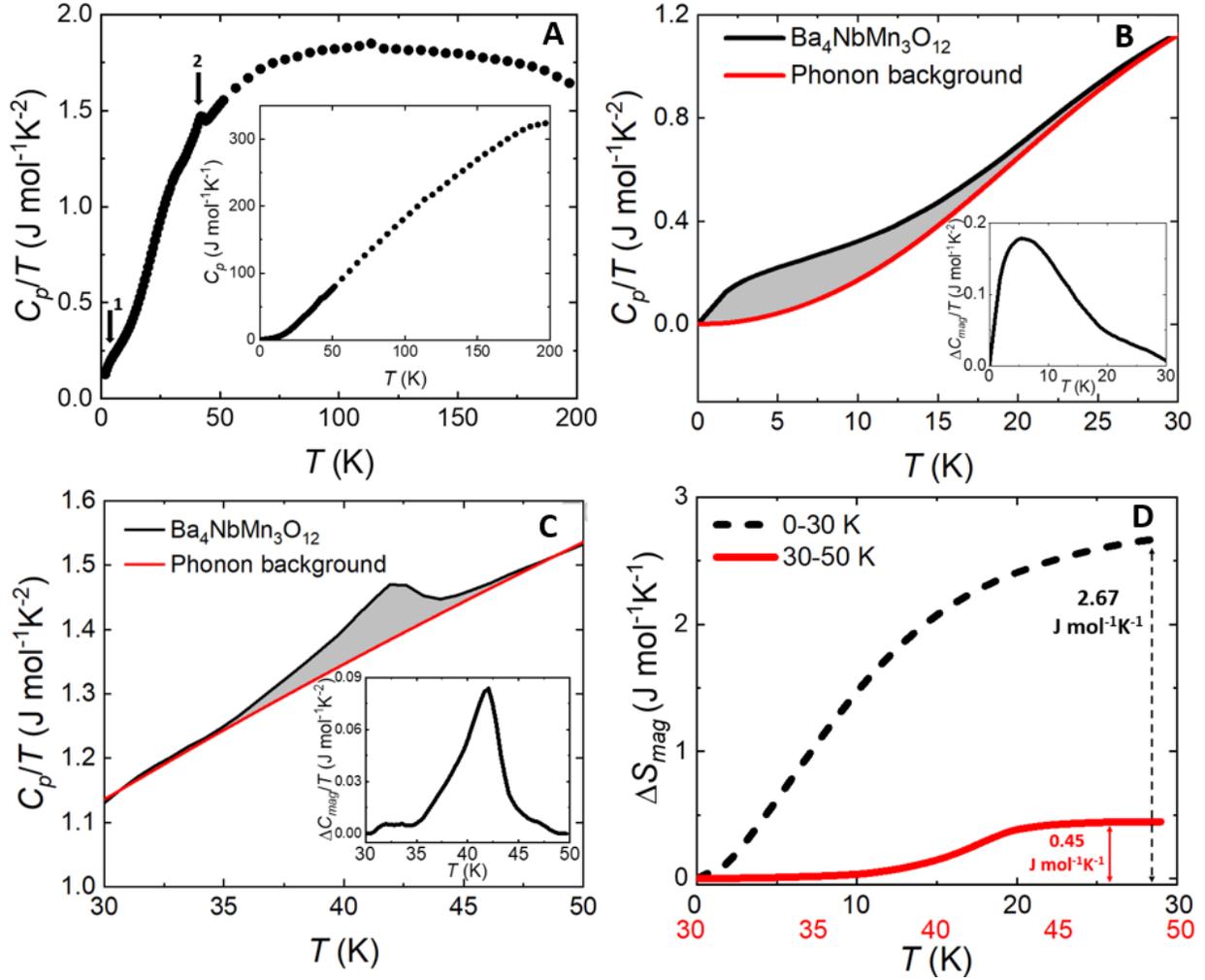

**Figure 7 (Color online):** (A) Molar heat capacity divided by temperature of Ba$_4$NbMn$_3$O$_{12}$ measured from 1.8 K to 200 K. Two transitions resulting in entropy losses are marked. A third one near 30 K is unmarked. Inset: the raw C$_p$ vs. T data. (B) The lower temperature transition showing a detail of the experimental data and an estimate of the phonon background. Inset: the C$_p$/T values in excess of the estimated phonon heat capacity. (C) The higher temperature transition showing a detail of the experimental data and an estimate of the phonon background. Inset the C$_p$/T values in excess of the estimated phonon heat capacity. (D) The entropy sums through the higher and lower temperature transitions, the former being 0.45 J mol$^{-1}$ K$^{-1}$ and the latter being 2.67 J mol$^{-1}$ K$^{-1}$.